\documentclass[preprint,showpacs,preprintnumbers,amsmath,amssymb,showkeys]{revtex4}

\usepackage{graphicx}
\usepackage{dcolumn}
\usepackage{bm}

\begin{document}


\title{
%
Field-induced charge transport at the surface of pentacene single crystals:
a method to study charge dynamics of 2D electron systems
in organic crystals
}

\author{J. Takeya$^{1,2}$}
\email{takeya@criepi.denken.or.jp}

\author{\ C. Goldmann$^1$}%
\author{S. Haas$^1$}
\author{K. P. Pernstich$^1$}
\author{B. Ketterer$^3$}
\author{B. Batlogg$^1$}
\affiliation{
$^1$Laboratory for Solid State Physics ETH, CH-8093
Z$\ddot{u}$rich, Switzerland
}%
\affiliation{
$^2$CRIEPI, 2-11-1, Iwado-kita, Komae, Tokyo 201-8511, Japan}
\affiliation{$^3$Laboratory for Micro- and Nanotechnology PSI, CH-5232
Villigen, Switzerland
}%

\date{\today}

\begin{abstract}
A method has been developed to inject {\it mobile} charges
at the surface of organic molecular crystals, and the
DC transport of field-induced holes has been measured at the surface
of pentacene single crystals.
To minimize damage to the soft and fragile surface,
the crystals are attached to a pre-fabricated substrate
which incorporates a gate dielectric (SiO$_2$) and
four probe pads.
The surface mobility of the pentacene crystals
ranges from 0.1 to 0.5 cm$^2$/Vs and
is nearly temperature-independent above $\sim 150$~K,
while it becomes thermally activated at lower temperatures
when the induced charges become localized.
Ruling out the influence of electric contacts and
crystal grain boundaries, the results contribute to the
microscopic understanding of trapping
and detrapping mechanisms in organic molecular crystals.

\end{abstract}

\maketitle


Transport of field-induced charge in crystals of
aromatic molecules is of interest
in both technological and academic contexts.
On one hand,
organic field-effect transistors (OFETs) provide key components
in \lq\lq plastic electronics" such as active matrix flexible displays,
where OFETs are used as pixel control devices \cite{application}.
On the other hand,
though a wide variety of organic materials have been studied,
relatively little is known about the details of
microscopic intermolecular-transport dynamics,
especially trapping and detrapping mechanisms,
even for simple molecules such as polyacenes \cite{pope}.
For example, a frequently addressed question is
how the field-effect mobility of OTFTs relates to molecular order
\cite{gundlach,dodabalpur}.
An experimental difficulty has been
that intrinsic transport properties are masked by structural
imperfections and the device performance is
limited by inefficient carrier injection because
most of the devices reported on to date involve
polycrystalline thin films.
Therefore, it is desirable to measure
field-induced transport in single crystals, applying a method
which avoids disturbing
the crystal surface as much as possible.
Of further scientific interest is a possible electronic phase transition to
intriguing states in such van-der-Waals crystals with moderate orbital overlaps
between adjacent molecules,
when the carrier density is high enough;
ordered electronic ground states have been demonstrated
in the form of
superconductivity in doped fullerenes and
complex charge-ordered states in charge-transfer compounds
\cite{c60,ctreview}.

Using the metal-oxide-semiconductor (MOS) structure,
one can continuously tune the amount of charge
confined to a two-dimensional layer near the surface,
so that the evolution
of the field-induced charge transport can be systematically studied.
However, the main difficulty lies in preventing the injected charge
from being trapped by interface states.
Although the surface of van-der-Waals crystals is
free from dangling bonds and can be highly ordered,
the surface can be damaged
when depositing a gate oxide film.
In this report, we present a general method of
fabricating molecular single crystal MOS structures
which avoids this problem:
a set of probing electrodes is
defined on top of a thermally oxidized
silicon wafer
which provides a gate insulator and
gate electrodes.
A separately grown pentacene
crystal is then attached on the substrate,
minimizing the damage to the crystal surface.
In such devices, the electrically induced charge is mobile due to
the good interface quality.
A crossover from thermally activated transport to temperature-independent
conduction is observed in
four-terminal measurements,
excluding parasitic effects from the source and drain contacts.

Figure 1(a) schematically shows the fabrication process of
the crystal-based MOS structure.
We use a heavily doped n-type silicon wafer
($p \simeq 0.016$~$\Omega$~cm @ RT) as a substrate.
The wafer is cleaned in a solution of sulfuric acid and hydrogen
peroxide @ 80~$^\circ$C,
then oxidized in dry O$_2$ @ 1150~$^\circ$C
to grow a high-quality gate insulator.
The surface roughness (as measured by AFM) over the
typical size of the device ($\sim 0.1$ mm)
is less than 1~nm, and the oxide has a thickness of 0.5-1~$\mu$m.
Probing electrodes
(1 nm Cr for adhesion and 9 nm Au)
for four-terminal measurements are defined by
electron-beam lithography.
The distance between the two voltage-probing pads is approx. 5~$\mu$m.
This small separation reduces the likelihood
of having macroscopic defects
such as molecular steps in the channel of the device.

Next, the SiO$_2$ is coated with
a self-assembled monolayer of octadecyltrichlorosilane (OTS)
with a thickness of $\sim$~3~nm.
It is found that the monolayer reduces
trapped charges at the interface \cite{gundlach1}.
To improve the electrical contact between
the gold pads and the molecular crystal, the substrate is
treated in a solution of 10 mM nitrobenzenethiol in ethanol.
The thiol bonds to the gold surface,
and the nitro group dopes
holes into the surface of the crystal
because of its large electron affinity.
The treatment significantly reduces the contact resistance
(approx. by a factor of 10 at room temperature and much more at low $T$),
which is crucial for
the low-temperature four-terminal measurement.
After the surface treatments,
the conductivity between the probe electrodes remains immeasurably small
when a gate voltage is applied, indicating that
the OTS monolayer is highly insulating.

Pentacene single crystals are grown separately by the physical
vapor transport technique \cite{vt}.
Starting from commercial powders of the same material (Fluka),
the growth is repeated three times to purify the crystals.
For a good adhesion to the substrate,
it is preferable to choose a thin transparent crystal, with
typical thickness of 2-5 $\mu$m.
Since the probing electrodes are visible through the crystal,
the crystal can be located
so that the most homogeneous part is placed between
the electrodes.
Electrostatic force ensures a strong adhesion of
the crystal to the substrate.
Alternatively, adhesion can be imposed by applying
a gate voltage.
A picture of a typical sample is shown in Fig.~1(b),
where no grain boundary is visible in the channel.

On these devices, four-terminal measurements have been
performed using four Keithley 6517A
electrometers/high-impedance meters.
Figure~1(c) shows the corresponding circuit diagram.
One of the electrometers applies the voltage $V_{GD}$
and monitors the leakage current $I_G$ through
the dielectric, another applies the source-drain voltage
and measures the current $I_S$ in the sample, and the others
measure the difference in potential between the two intermediate electrodes
and the drain electrode ($V_1$ and $V_2$).
The sheet conductivity $\sigma_\Box$ of the channel is given by
$I_S / (V_2-V_1) \cdot l / w$, regardless of the
voltage drops across the source and drain contacts ($l$ and $w$ are
the distance between the two intermediate electrodes
and the width of the channel, respectively).
The drain electrode is set common for the $V_{GD}$ and $V_{SD}$
application (in contrary to the standard connection
of MOS transistors).
Since only the \lq\lq linear regime" of the channel
$I-V$ characteristics is of interest for the
four-terminal method,
the polarity of $V_{SD}$ is chosen opposite to the polarity
of $V_{GD}$
to avoid the \lq\lq pinch-off" saturation in $I_S$.
For convenience, we additionally define
$V_G \equiv V_{GD} - ( V_1 + V_2 ) / 2$,
which represents an effective gate voltage
with respect to the central part of the channel
where the sheet conductivity is measured.

The distance between the two current-injecting electrodes is approximately
10-15 times longer than the width of these electrodes,
so that the most of the current flows through the field-accumulated channel
between them.
In order to estimate the influence of the
current flowing via the voltage electrodes, we also performed
measurement
using the circuitry illustrated in Figure~1(d);
using a guarding function with a high input impedance (a voltage follower)
of the Keithley 6517A,
a set of two voltage pads (no. 1 and 3 in the figure)
is held at the same potential level,
so that no current flows between these pads.
The channel conductivity in the guarded circuit was measured to be
$\sim 6$\% smaller than in the unguarded circuit.

A drawback of the guarded measurement is its slower response,
as it is necessary to minimize
the redistribution of trap centers at the interface
in response to high electric fields applied over a long time.
As reported by Schoonveld {\it et al.}
for pentacene thin-film devices
on SiO$_2$ \cite{schoonveld1},
localized charges randomly distributed
in pentacene and/or the SiO$_2$ can
slowly move towards the interface
in response to a relatively high gate voltage.
Therefore,
the distribution of the hole-trapping centers
changes and the measured values of the field-induced current
are not exactly reproduced after prolonged influence of $V_G$.
The redistribution is faster at higher temperatures and at
higher gate voltages.
In order to minimize the effect of this charge relaxation,
we swept $V_{GD}$ at a fixed $V_{SD}$ as quickly as $\sim 30$~s
per sweep, using the unguarded circuitry,
minimizing hysteresis effects.
Moreover, $I_{SD}$ thus measured is reproduced
after a set of measurements at different temperatures,
ensuring that the relaxation effect is
negligible in the measurements.

In Fig.~2(a) and (b), $I_S$ at fixed values of $V_{SD}$ is
plotted for two temperatures as a function of $V_G$.
The source current in the pentacene single crystal
substantially increases due to hole injection by the negative $V_G$.
On the other hand, no enhancement of $I_S$ is detected
with positive gate voltage.
The influence of
leakage current through the dielectric $I_G$ is
negligibly small (less than 0.1\% of $I_S$).
In Fig.~2(c), $\sigma_\Box$ is plotted against $V_G$,
where 6\% is subtracted
from the result of the unguarded measurement.
The curves for different values of $V_{SD}$ are close to each other,
indicating a nearly ohmic current-voltage relationship for the
field-induced conductivity at the surface \cite{nonohmic}.
The observed curvature resembles that of typical organic
field-effect transistors (FETs) and is consistent with
a standard model:
the injected charge is trapped until $V_G$ reaches a threshold voltage
$V_{th}$, and
the amount of trap-free charge is given by
$p_{free} e = \epsilon \epsilon_0 (V_G - V_{th}) / d$.
In reality, the crossover is usually gradual \cite{horovitz,horovitz_rev}.
As the surface mobility $\mu$ equals to $\sigma_\Box / (p_{free} e)$,
where $e$ is the charge of an electron,
$\mu$ for this sample can be estimated from the slope
to be $\sim$~0.1~cm$^2$/Vs with an uncertainty of 15-20\%.

Figure~3(a) presents the $\sigma_\Box$-$V_G$ curves at different temperatures
from 160~K to 260~K.
While the temperature dependence is obvious in the low $V_G$ region,
the maximum slope
at high $V_G$ does not differ much as a function of temperature.
For each temperature, the mobility is estimated from this maximum slope
and is plotted in Fig.~3(b).
The results for two more crystals are also shown.
The mobility values
range from 0.1 to 0.5~cm$^2$/Vs, with no apparent temperature dependence.
These mobilities are comparable to typical room temperature bulk values
measured in naphthalene by the time-of-flight method \cite{schein},
though an ultraclean crystal can show a rapid increase in $\mu(T)$ upon
cooling \cite{warta,karl}.
The mobility values of our devices are also comparable to those of
high-quality pentacene thin-film FETs, though somewhat smaller
than the best MOS device (1.5~cm$^2$/Vs) \cite{nelson}.
(More recently, a mobility of
3~cm$^2$/Vs was reported for pentacene OTFTs with
a polymeric dielectric \cite{klauk}.)
The absence of a temperature dependence is reported for
the high-mobility devices of Ref.~\onlinecite{nelson} as well;
however, possible contributions
from grain boundaries and electric contacts
limited further discussion of the origin of the
$T$ independence.
Since it is not likely that such extrinsic mechanisms dominate $\mu(T)$
in our four-terminal measurements of single crystals,
the nearly temperature independent
$\mu(T)$ (at least from $\sim$~150~K to near room temperature) appears
to be an intrinsic property of the surface
of the pentacene single crystals in this study,
motivating us to investigate the fundamental
inter-molecular charge dynamics more quantitatively.

The nearly $T$-independent feature is inconsistent with
usual hopping models which predict strong $T$ dependence
as $\sim T^{-m} exp(-\Delta / k_B T)$.
A difficulty arises in the band picture as well
because a naive estimation of the mean free path $\ell$
results in a value shorter than
the distance between adjacent molecules $a \sim 5$~\AA;
assuming the free-electron approximation
$\mu = e \tau / m^* = e \ell / m^* \overline{v}$
($\tau$ and $m^*$ are the relaxation time and the effective mass,
respectively) and the 2D Boltzmann distribution of the average velocity
$\overline{v} = \sqrt{2 k_B T / m^*}$,
$\ell / a$ is evaluated as $\sim $~0.1,
even if $\mu \sim 0.5$~cm$^2$/Vs, $T = 300$~K,
and $m^*$ is 1.5-5 times
the free-electron mass \cite{bandcal}
(the Fermi temperature is as low as $\sim 40$~K even at $V_G = 100$~V).
Allowing for polaron formation,
which is common in dilute-charge systems,
the nearly $T$-independent feature is deduced
for naphthalene \cite{kenkre};
however, since the calculation is based on
a balance between electronic
parameters (such as band width) and phononic
parameters,
the applicability to the case of pentacene crystals
would need to be further explored.
The combination of a short $\ell$ and an almost $T$-independent
conductivity is reminiscent of an almost $T$ independent
inter-layer charge transport near room temperature
in a high-$T_c$ cuprate.
There,
metallic planes are stacked in a sub-nanometer distance.
It is argued that direct tunneling
between the adjacent planes dominates the inter-layer transport
in such systems \cite{bi}.

In order to gain more insight into the
transport mechanism of the field-induced charges,
we extended the measurements on Sample C to
lower temperatures,
where the influence of the traps remains visible
even at the highest applied gate voltage.
The field-induced conductivity at fixed gate voltages
$\sigma_\Box^{FET} (V_G)$ ($\equiv \sigma_\Box (V_G) - \sigma_\Box (0)$)
is plotted in Fig.~4 as a function of
temperature.
The amount of induced charge $p$ shown in the figure
is calculated from $V_G$.
The conductivity is nearly temperature independent above $\sim 150$~K
reflecting $\mu(T)$ shown in Fig.~3(b),
and rapidly diminishes with an activation-like temperature dependence
upon further cooling.
The \lq\lq activation energy" $\Delta$, estimated from
the slope of the $\log \sigma_\Box^{FET} - 1 / T$ plot,
is comparable to or less than room temperature and
slightly decreases with increasing $V_G$ (inset Fig.~4).
The observation on the low-$T$ side
is similar to recent results
by Schoonveld {\it et al.} on a large-grain thin-film FET
\cite{schoonveld},
where the channel is restricted to a single domain
and the influence of the contacts is analytically subtracted.
The value of $\Delta$ in ref.~\onlinecite{schoonveld}
is similar to the ones we report here, and smaller
at higher $V_G$, though the room temperature conductivity
(and mobility) is more than one order smaller than that of Sample C.
Therefore, the observed temperature and $V_G$ dependence may
reflect the generic behavior of the
the conductivity at the surface of organic crystals
influenced by shallow trap potentials.
The dependence of $\Delta$ on $V_G$ can be rationalized
assuming a distribution of the trap levels as in amorphous semiconductors
\cite{amorphous},
because an increasing number of injected holes lowers
the energy necessary to activate localized holes up to
the \lq\lq mobility edge".
However,
one would question the reason of such qualitative similarity
between the transport in single crystals and in amorphous systems,
taking into consideration
the extreme difference in the extent of the molecular order.
In this sense, the microscopic mechanism causing the localization
still remains to be elucidated
in organic field-effect transistors.

An intriguing mechanism is proposed
in ref.~\onlinecite{schoonveld} for the charge localization;
random charges located at defects near the interface induce
\lq\lq offset charges" in the molecules,
so that the field-induced holes must hop from molecule
to molecule feeling a random potential
via the charging energy due to the Coulomb interaction.
A noteworthy prediction is that a $T$-independent tunneling
probability governs the inter-molecular charge transport
when the above localization mechanism becomes less significant
because of either high temperature or
high average hopping rate between the molecules.
This may provide an interpretation of an almost $T$-independent
$\mu$ as intrinsic to high-mobility pentacene field-effect devices.
Measurements on single-crystal devices of various
molecular compounds may help to elucidate the microscopic mechanism
limiting the field-induced charge transport.
Very recently, single-crystal measurements
have indeed been successful
\cite{podzorov,butko,hasegawa,morpurgo}.

In summary,
we have studied the transport of field-induced charge in pentacene
single crystals by means of a new four-terminal method
in which extrinsic contributions from grain boundaries
and electric contacts are minimized.
The hole mobility is comparable to the one in
high-quality pentacene OTFTs.
The induced sheet carrier densities
are of the order from $10^{11}$ to $10^{12}$~cm$^{-2}$
and lead to a temperature
independent conductivity above $\sim$~150~K which correlates well with a high
mobility of $\sim$~0.5~cm$^2$/Vs.
At lower temperatures, a crossover to
thermally activated transport governed by shallow traps is observed.
The presented method
is well suited to investigate transport in
other organic crystals and to study
fundamental charge transport mechanisms in the 2D electronic systems
in molecular crystals.

The authors acknowledge helpful discussions with D. Gundlach,
T. Hasegawa, Ch. Bergemann, K. Ensslin, Ch. Helm, S. Wehrli,
M. Sigrist and R. Kleiner.
We also thank K. Mattenberger, H.-P. Staub, M. Moessle,
V. Oehmichen, H. Scherrer, N. Boos,
S. Kicin, A. Fuhrer, and Ch. Ellenberger for technical support.
This study was supported by the Swiss National Science Foundation.

\pagebreak

\bibliography{penfe16}

\begin{thebibliography}{26}
\expandafter\ifx\csname natexlab\endcsname\relax\def\natexlab#1{#1}\fi
\expandafter\ifx\csname bibnamefont\endcsname\relax
  \def\bibnamefont#1{#1}\fi
\expandafter\ifx\csname bibfnamefont\endcsname\relax
  \def\bibfnamefont#1{#1}\fi
\expandafter\ifx\csname citenamefont\endcsname\relax
  \def\citenamefont#1{#1}\fi
\expandafter\ifx\csname url\endcsname\relax
  \def\url#1{\texttt{#1}}\fi
\expandafter\ifx\csname urlprefix\endcsname\relax\def\urlprefix{URL }\fi
\providecommand{\bibinfo}[2]{#2}
\providecommand{\eprint}[2][]{\url{#2}}

\bibitem[{\citenamefont{Dimitrakopoulos and Malenfant}(2002)}]{application}
\bibinfo{author}{\bibfnamefont{C.}~\bibnamefont{Dimitrakopoulos}}
  \bibnamefont{and}
  \bibinfo{author}{\bibfnamefont{P.}~\bibnamefont{Malenfant}},
  \bibinfo{journal}{Adv.\ Mater.} \textbf{\bibinfo{volume}{14}},
  \bibinfo{pages}{99} (\bibinfo{year}{2002}).

\bibitem[{\citenamefont{Pope and Swenberg}(1999)}]{pope}
\bibinfo{author}{\bibfnamefont{M.}~\bibnamefont{Pope}} \bibnamefont{and}
  \bibinfo{author}{\bibfnamefont{C.}~\bibnamefont{Swenberg}},
  \emph{\bibinfo{title}{Electronic Processes in Organic Crystals and Polymers,
  2nd ed.}} (\bibinfo{publisher}{Oxford University Press},
  \bibinfo{year}{1999}).

\bibitem[{\citenamefont{Gundlach et~al.}(1997)\citenamefont{Gundlach, Lin,
  T.N.Jackson, Nelson, and Schlom}}]{gundlach}
\bibinfo{author}{\bibfnamefont{D.~J.} \bibnamefont{Gundlach}},
  \bibinfo{author}{\bibfnamefont{Y.~Y.} \bibnamefont{Lin}},
  \bibinfo{author}{\bibnamefont{T.N.Jackson}},
  \bibinfo{author}{\bibfnamefont{S.~F.} \bibnamefont{Nelson}},
  \bibnamefont{and} \bibinfo{author}{\bibfnamefont{D.~G.}
  \bibnamefont{Schlom}}, \bibinfo{journal}{IEEE Elect. Dev. Lett.}
  \textbf{\bibinfo{volume}{18}}, \bibinfo{pages}{87} (\bibinfo{year}{1997}).

\bibitem[{\citenamefont{Dodabalpur et~al.}(1995)\citenamefont{Dodabalpur,
  Torsi, and Katz}}]{dodabalpur}
\bibinfo{author}{\bibfnamefont{A.}~\bibnamefont{Dodabalpur}},
  \bibinfo{author}{\bibfnamefont{L.}~\bibnamefont{Torsi}}, \bibnamefont{and}
  \bibinfo{author}{\bibfnamefont{H.~E.} \bibnamefont{Katz}},
  \bibinfo{journal}{Science} \textbf{\bibinfo{volume}{268}},
  \bibinfo{pages}{270} (\bibinfo{year}{1995}).

\bibitem[{\citenamefont{Dresselhaus et~al.}(1996)\citenamefont{Dresselhaus,
  Dresselhaus, and P.C.Eklund}}]{c60}
\bibinfo{author}{\bibfnamefont{M.}~\bibnamefont{Dresselhaus}},
  \bibinfo{author}{\bibfnamefont{G.}~\bibnamefont{Dresselhaus}},
  \bibnamefont{and} \bibinfo{author}{\bibnamefont{P.C.Eklund}},
  \emph{\bibinfo{title}{Science of Fullerenes and Carbon Nanotubes}}
  (\bibinfo{publisher}{Academic Press}, \bibinfo{year}{1996}).

\bibitem[{\citenamefont{Ishiguro et~al.}(1997)\citenamefont{Ishiguro, Yamaji,
  and Saito}}]{ctreview}
\bibinfo{author}{\bibfnamefont{T.}~\bibnamefont{Ishiguro}},
  \bibinfo{author}{\bibfnamefont{K.}~\bibnamefont{Yamaji}}, \bibnamefont{and}
  \bibinfo{author}{\bibfnamefont{G.}~\bibnamefont{Saito}},
  \emph{\bibinfo{title}{Organic Superconductors, 2nd ed.}}
  (\bibinfo{publisher}{Springer, Berlin}, \bibinfo{year}{1997}).

\bibitem[{\citenamefont{Gundlach et~al.}(2001)\citenamefont{Gundlach, Jia, and
  T.N.Jackson}}]{gundlach1}
\bibinfo{author}{\bibfnamefont{D.~J.} \bibnamefont{Gundlach}},
  \bibinfo{author}{\bibfnamefont{L.-L.} \bibnamefont{Jia}}, \bibnamefont{and}
  \bibinfo{author}{\bibnamefont{T.N.Jackson}}, \bibinfo{journal}{IEEE Electr.
  Dev. Lett.} \textbf{\bibinfo{volume}{22}}, \bibinfo{pages}{571}
  (\bibinfo{year}{2001}).

\bibitem[{\citenamefont{Kloc et~al.}(1997)\citenamefont{Kloc, Simpkins,
  Siegrist, and Laudise}}]{vt}
\bibinfo{author}{\bibfnamefont{C.}~\bibnamefont{Kloc}},
  \bibinfo{author}{\bibfnamefont{P.~G.} \bibnamefont{Simpkins}},
  \bibinfo{author}{\bibfnamefont{T.}~\bibnamefont{Siegrist}}, \bibnamefont{and}
  \bibinfo{author}{\bibfnamefont{R.~A.} \bibnamefont{Laudise}},
  \bibinfo{journal}{J.\ Cryst.\ Growth} \textbf{\bibinfo{volume}{182}},
  \bibinfo{pages}{416} (\bibinfo{year}{1997}).

\bibitem[{\citenamefont{Schoonveld et~al.}(1998)\citenamefont{Schoonveld,
  Vrijmoeth, and Klapwijk}}]{schoonveld1}
\bibinfo{author}{\bibfnamefont{W.~A.} \bibnamefont{Schoonveld}},
  \bibinfo{author}{\bibfnamefont{J.}~\bibnamefont{Vrijmoeth}},
  \bibnamefont{and} \bibinfo{author}{\bibfnamefont{T.~M.}
  \bibnamefont{Klapwijk}}, \bibinfo{journal}{Appl.\ Phys.\ Lett.}
  \textbf{\bibinfo{volume}{73}}, \bibinfo{pages}{3884} (\bibinfo{year}{1998}).

\bibitem[{non(a)}]{nonohmic}
\emph{\bibinfo{title}{a}} (\bibinfo{year}{a}), \bibinfo{note}{a. Too high
  $V_{SD}$ can detrap some localized charges and may introduce nonohmicity. In
  this report, we restrict $V_{SD}$ to the nearly ohmic regime to avoid
  complication.}

\bibitem[{\citenamefont{Horovitz et~al.}(2000)\citenamefont{Horovitz, Hajlaoui,
  and Hajlaoui}}]{horovitz}
\bibinfo{author}{\bibfnamefont{G.}~\bibnamefont{Horovitz}},
  \bibinfo{author}{\bibfnamefont{M.~E.} \bibnamefont{Hajlaoui}},
  \bibnamefont{and} \bibinfo{author}{\bibfnamefont{R.}~\bibnamefont{Hajlaoui}},
  \bibinfo{journal}{J.\ Appl.\ Phys.} \textbf{\bibinfo{volume}{87}},
  \bibinfo{pages}{4456} (\bibinfo{year}{2000}).

\bibitem[{\citenamefont{Horovitz}(1998)}]{horovitz_rev}
\bibinfo{author}{\bibfnamefont{G.}~\bibnamefont{Horovitz}},
  \bibinfo{journal}{Adv.\ Mater.} \textbf{\bibinfo{volume}{10}},
  \bibinfo{pages}{365} (\bibinfo{year}{1998}).

\bibitem[{\citenamefont{Schein and McGie}(1979)}]{schein}
\bibinfo{author}{\bibfnamefont{L.~B.} \bibnamefont{Schein}} \bibnamefont{and}
  \bibinfo{author}{\bibfnamefont{A.~R.} \bibnamefont{McGie}},
  \bibinfo{journal}{Phys.\ Rev.\ B} \textbf{\bibinfo{volume}{20}},
  \bibinfo{pages}{1631} (\bibinfo{year}{1979}).

\bibitem[{\citenamefont{Warta et~al.}(1985)\citenamefont{Warta, Stehle, and
  Karl}}]{warta}
\bibinfo{author}{\bibfnamefont{W.}~\bibnamefont{Warta}},
  \bibinfo{author}{\bibfnamefont{R.}~\bibnamefont{Stehle}}, \bibnamefont{and}
  \bibinfo{author}{\bibfnamefont{N.}~\bibnamefont{Karl}},
  \bibinfo{journal}{Appl.\ Phys.\ A} \textbf{\bibinfo{volume}{36}},
  \bibinfo{pages}{163} (\bibinfo{year}{1985}).

\bibitem[{\citenamefont{Karl}(2001)}]{karl}
\bibinfo{author}{\bibfnamefont{N.}~\bibnamefont{Karl}},
  \emph{\bibinfo{title}{in Organic Electronic Materials: conjugated polymers
  and low molecular weight organic solids, ed. by R. Farchioni and G. Grosso}}
  (\bibinfo{publisher}{Springer}, \bibinfo{year}{2001}).

\bibitem[{\citenamefont{Nelson et~al.}(1998)\citenamefont{Nelson, Lin,
  Gundlach, and Jackson}}]{nelson}
\bibinfo{author}{\bibfnamefont{S.~F.} \bibnamefont{Nelson}},
  \bibinfo{author}{\bibfnamefont{Y.-Y.} \bibnamefont{Lin}},
  \bibinfo{author}{\bibfnamefont{D.~J.} \bibnamefont{Gundlach}},
  \bibnamefont{and} \bibinfo{author}{\bibfnamefont{T.~N.}
  \bibnamefont{Jackson}}, \bibinfo{journal}{Appl.\ Phys.\ Lett.}
  \textbf{\bibinfo{volume}{72}}, \bibinfo{pages}{1854} (\bibinfo{year}{1998}).

\bibitem[{\citenamefont{Klauk et~al.}(2002)\citenamefont{Klauk, Halik,
  Zschieschang, Schmid, Radlik, and Weber}}]{klauk}
\bibinfo{author}{\bibfnamefont{H.}~\bibnamefont{Klauk}},
  \bibinfo{author}{\bibfnamefont{M.}~\bibnamefont{Halik}},
  \bibinfo{author}{\bibfnamefont{U.}~\bibnamefont{Zschieschang}},
  \bibinfo{author}{\bibfnamefont{G.}~\bibnamefont{Schmid}},
  \bibinfo{author}{\bibfnamefont{W.}~\bibnamefont{Radlik}}, \bibnamefont{and}
  \bibinfo{author}{\bibfnamefont{W.}~\bibnamefont{Weber}},
  \bibinfo{journal}{J.\ Appl.\ Phys.} \textbf{\bibinfo{volume}{92}},
  \bibinfo{pages}{5259} (\bibinfo{year}{2002}).

\bibitem[{\citenamefont{Cornil et~al.}(2001)\citenamefont{Cornil, Calbert, and
  Br$\acute{e}$das}}]{bandcal}
\bibinfo{author}{\bibfnamefont{J.}~\bibnamefont{Cornil}},
  \bibinfo{author}{\bibfnamefont{J.~P.} \bibnamefont{Calbert}},
  \bibnamefont{and} \bibinfo{author}{\bibfnamefont{J.~L.}
  \bibnamefont{Br$\acute{e}$das}}, \bibinfo{journal}{J. Am. Chem. Soc.}
  \textbf{\bibinfo{volume}{123}}, \bibinfo{pages}{1250} (\bibinfo{year}{2001}),
  \bibinfo{note}{rR. C. Haddon, X. Chi, M. E. Itkis, J. E. Anthony, D. L.
  Eaton, T. Siegrist, C. C. Mattheus, and T. T. M. Palstra, J. Phys. Chem. B.
  \bf{106} \rm{, 8288 (2002),} M. L. Tiago, J. E. Northrup, and S. G. Louie,
  Phys. Rev. B \bf{67} \rm{, 115212 (2003)}.}

\bibitem[{\citenamefont{Kenkre et~al.}(1989)\citenamefont{Kenkre, Andersen,
  Dunlap, and Duke}}]{kenkre}
\bibinfo{author}{\bibfnamefont{V.~M.} \bibnamefont{Kenkre}},
  \bibinfo{author}{\bibfnamefont{J.~D.} \bibnamefont{Andersen}},
  \bibinfo{author}{\bibfnamefont{D.~H.} \bibnamefont{Dunlap}},
  \bibnamefont{and} \bibinfo{author}{\bibfnamefont{C.~B.} \bibnamefont{Duke}},
  \bibinfo{journal}{Phys.\ Rev.\ Lett.} \textbf{\bibinfo{volume}{62}},
  \bibinfo{pages}{1165} (\bibinfo{year}{1989}).

\bibitem[{\citenamefont{Cooper and Grey}(1994)}]{bi}
\bibinfo{author}{\bibfnamefont{S.~L.} \bibnamefont{Cooper}} \bibnamefont{and}
  \bibinfo{author}{\bibfnamefont{K.~E.} \bibnamefont{Grey}},
  \emph{\bibinfo{title}{in Physical Properties of High Temperature
  Superconductors, ed. by D. M. Ginsberg}}, vol.~\bibinfo{volume}{IV}
  (\bibinfo{publisher}{World Scientific}, \bibinfo{year}{1994}).

\bibitem[{\citenamefont{Schoonveld et~al.}(2000)\citenamefont{Schoonveld,
  Wildeman, Fichou, Bobbert, van Wees, and Klapwijk}}]{schoonveld}
\bibinfo{author}{\bibfnamefont{W.~A.} \bibnamefont{Schoonveld}},
  \bibinfo{author}{\bibfnamefont{J.}~\bibnamefont{Wildeman}},
  \bibinfo{author}{\bibfnamefont{D.}~\bibnamefont{Fichou}},
  \bibinfo{author}{\bibfnamefont{P.~A.} \bibnamefont{Bobbert}},
  \bibinfo{author}{\bibfnamefont{B.~J.} \bibnamefont{van Wees}},
  \bibnamefont{and} \bibinfo{author}{\bibfnamefont{T.~M.}
  \bibnamefont{Klapwijk}}, \bibinfo{journal}{Nature}
  \textbf{\bibinfo{volume}{404}}, \bibinfo{pages}{977} (\bibinfo{year}{2000}).

\bibitem[{\citenamefont{Vissenberg and Matters}(1998)}]{amorphous}
\bibinfo{author}{\bibfnamefont{M.~C. J.~M.} \bibnamefont{Vissenberg}}
  \bibnamefont{and} \bibinfo{author}{\bibfnamefont{M.}~\bibnamefont{Matters}},
  \bibinfo{journal}{Phys.\ Rev.\ B} \textbf{\bibinfo{volume}{57}},
  \bibinfo{pages}{12964} (\bibinfo{year}{1998}).

\bibitem[{\citenamefont{Podzorov et~al.}(2003)\citenamefont{Podzorov, Pudalov,
  and Gershenson}}]{podzorov}
\bibinfo{author}{\bibfnamefont{V.}~\bibnamefont{Podzorov}},
  \bibinfo{author}{\bibfnamefont{V.~M.} \bibnamefont{Pudalov}},
  \bibnamefont{and} \bibinfo{author}{\bibfnamefont{M.~E.}
  \bibnamefont{Gershenson}}, \bibinfo{journal}{Appl. Phys. Lett.}
  \textbf{\bibinfo{volume}{82}}, \bibinfo{pages}{1739} (\bibinfo{year}{2003}).

\bibitem[{\citenamefont{Butko et~al.}(2003)\citenamefont{Butko, Chi, Lang, and
  Ramirez}}]{butko}
\bibinfo{author}{\bibfnamefont{V.~Y.} \bibnamefont{Butko}},
  \bibinfo{author}{\bibfnamefont{X.}~\bibnamefont{Chi}},
  \bibinfo{author}{\bibfnamefont{D.~V.} \bibnamefont{Lang}}, \bibnamefont{and}
  \bibinfo{author}{\bibfnamefont{A.~P.} \bibnamefont{Ramirez}},
  \bibinfo{journal}{cond-mat/0305402}  (\bibinfo{year}{2003}).

\bibitem[{\citenamefont{Hasegawa et~al.}(2003)\citenamefont{Hasegawa,
  Mattenberger, Takeya, and Batlogg}}]{hasegawa}
\bibinfo{author}{\bibfnamefont{T.}~\bibnamefont{Hasegawa}},
  \bibinfo{author}{\bibfnamefont{K.}~\bibnamefont{Mattenberger}},
  \bibinfo{author}{\bibfnamefont{J.}~\bibnamefont{Takeya}}, \bibnamefont{and}
  \bibinfo{author}{\bibfnamefont{B.}~\bibnamefont{Batlogg}},
  \bibinfo{journal}{preprint}  (\bibinfo{year}{2003}).

\bibitem[{\citenamefont{Morpurgo}(2003)}]{morpurgo}
\bibinfo{author}{\bibfnamefont{A.}~\bibnamefont{Morpurgo}},
  \bibinfo{journal}{private communication}  (\bibinfo{year}{2003}).

\end{thebibliography}

\pagebreak

\begin{figure}
\begin{flushleft}
Figure captions:
\end{flushleft}
\caption{\label{fig:epsart} (a) Fabrication of the field-effect device
with an organic single crystal MOS structure.
The crystal is attached to the substrate in the end of the fabrication
process.
(b) Top view of a typical sample with a pentacene single crystal.
The gold electrodes are visible through the transparent crystal.
(c) and (d) Circuit diagrams
to measure the transport of field-induced charge
by means of the four-terminal method.  (d) incorporates a guarding circuit to
cut the extra current path through the voltage pads.
}

\caption{\label{fig:epsart2} Four-probe measurement of a pentacene
single crystal field-effect device. Shown is the dependence
of three quantities on the gate voltage $V_G$
(a) the source current $I_S$ at 260~K,
(b) $I_S$ at 180~K,  and
(c) the channel conductivity at both temperatures.
The curves in (c) are represented with the same markers as in (a) and (b)
for corresponding source-drain voltages.
}

\caption{\label{fig:epsart3} (a) Channel conductivity vs. gate voltage
at different temperatures from 160~K to 260~K.
The field-effect mobility is derived from
the slope of the solid lines.
(b) The field-effect mobility as a function of temperature
for three different pentacene single crystal devices.}

\caption{\label{fig:epsart4} Main panel: The field-induced conductivity
(logarithmic scale) at four different gate voltages
as a function of inverse temperature (Sample C).
Dashed lines are a guide to the eye to illustrate
the crossover from nearly $T$ independent to activation like $T$ dependence.
Inset: Activation energy, estimated from the slope in the low-$T$
regime, as function of estimated total hole density.
}
\end{figure}


\begin{figure}
\includegraphics[width=14cm]{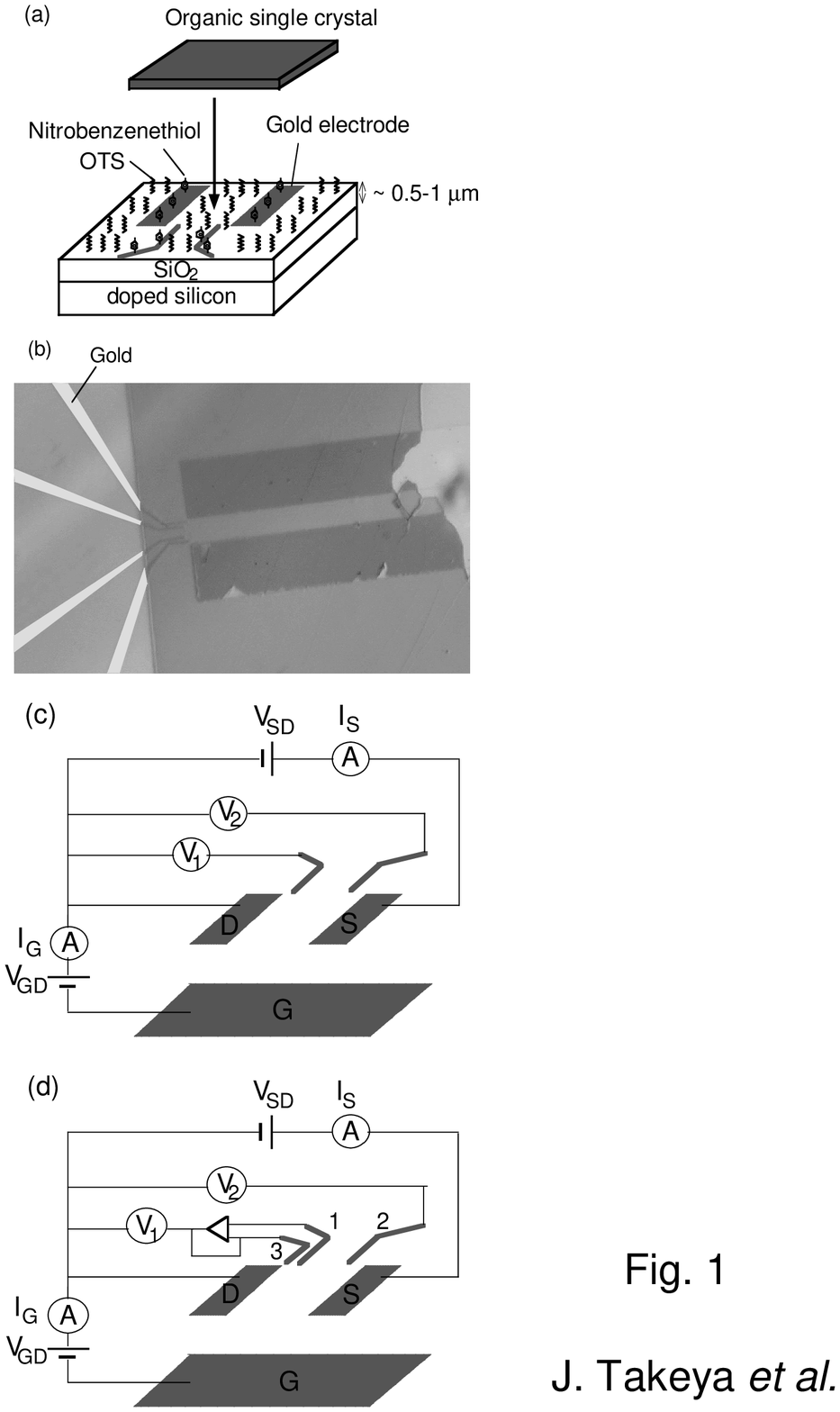}
\end{figure}

\begin{figure}
\includegraphics[width=17cm]{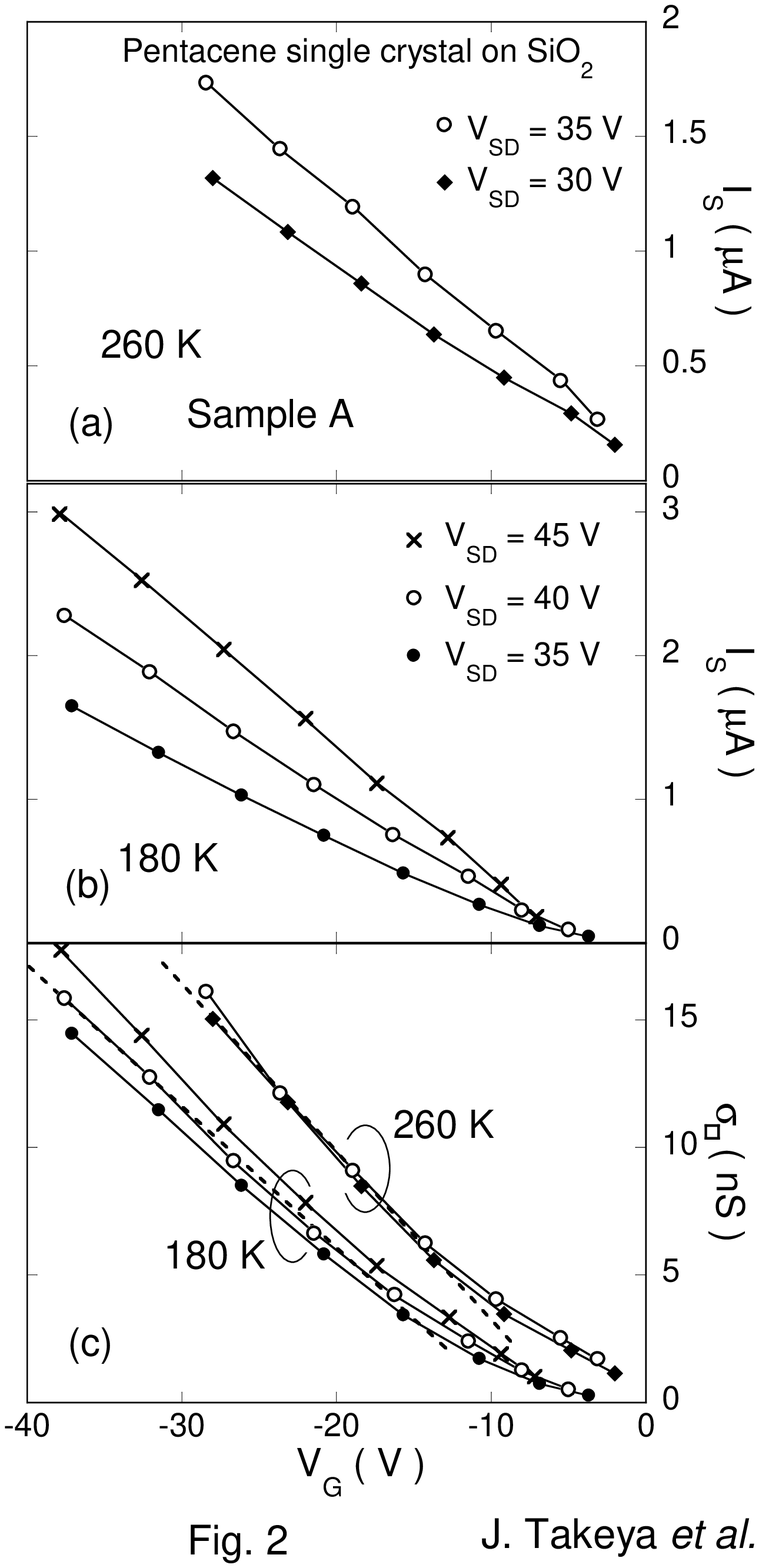}
\end{figure}

\begin{figure}
\includegraphics[width=17cm]{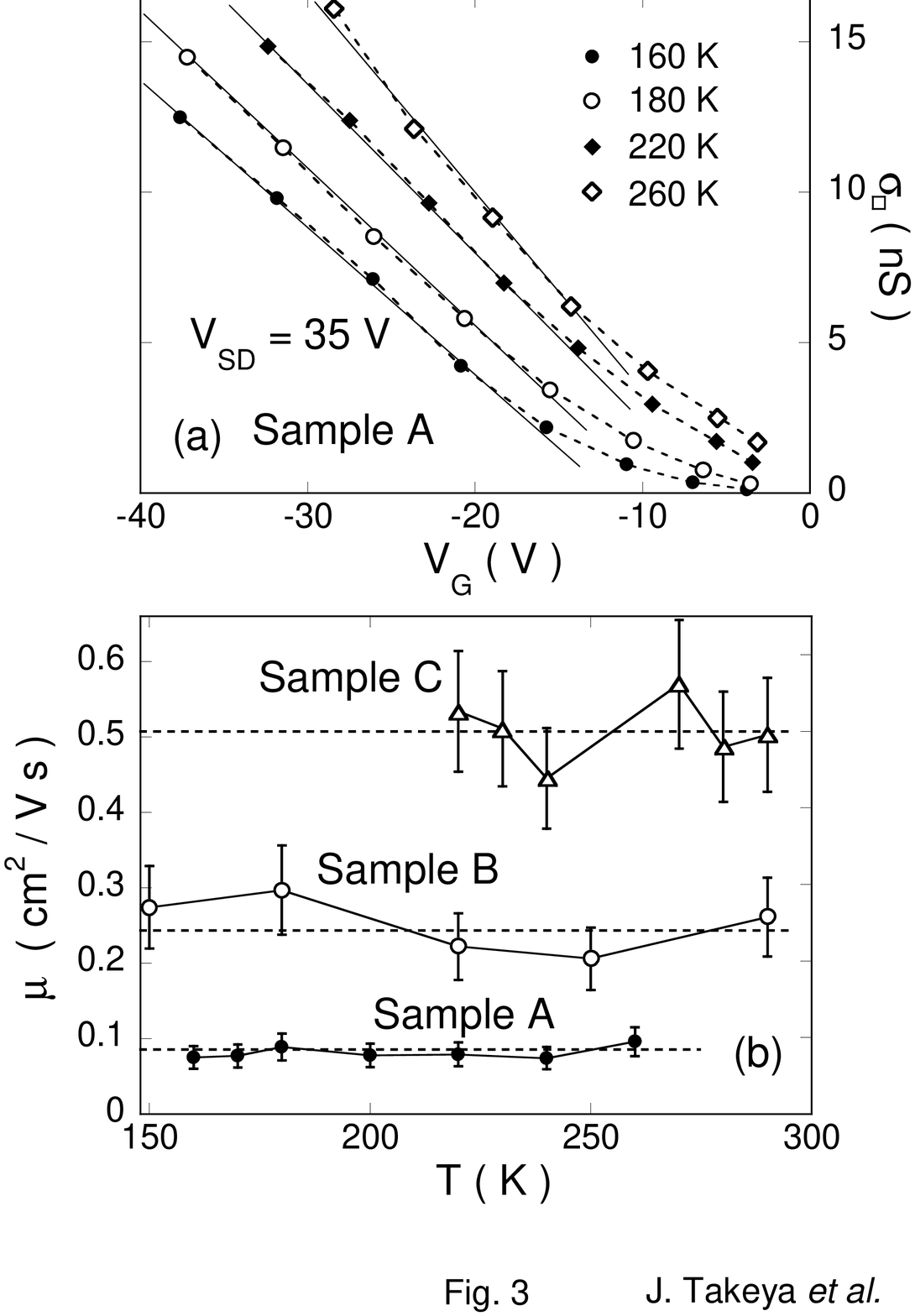}
\end{figure}

\begin{figure}
\includegraphics[width=17cm]{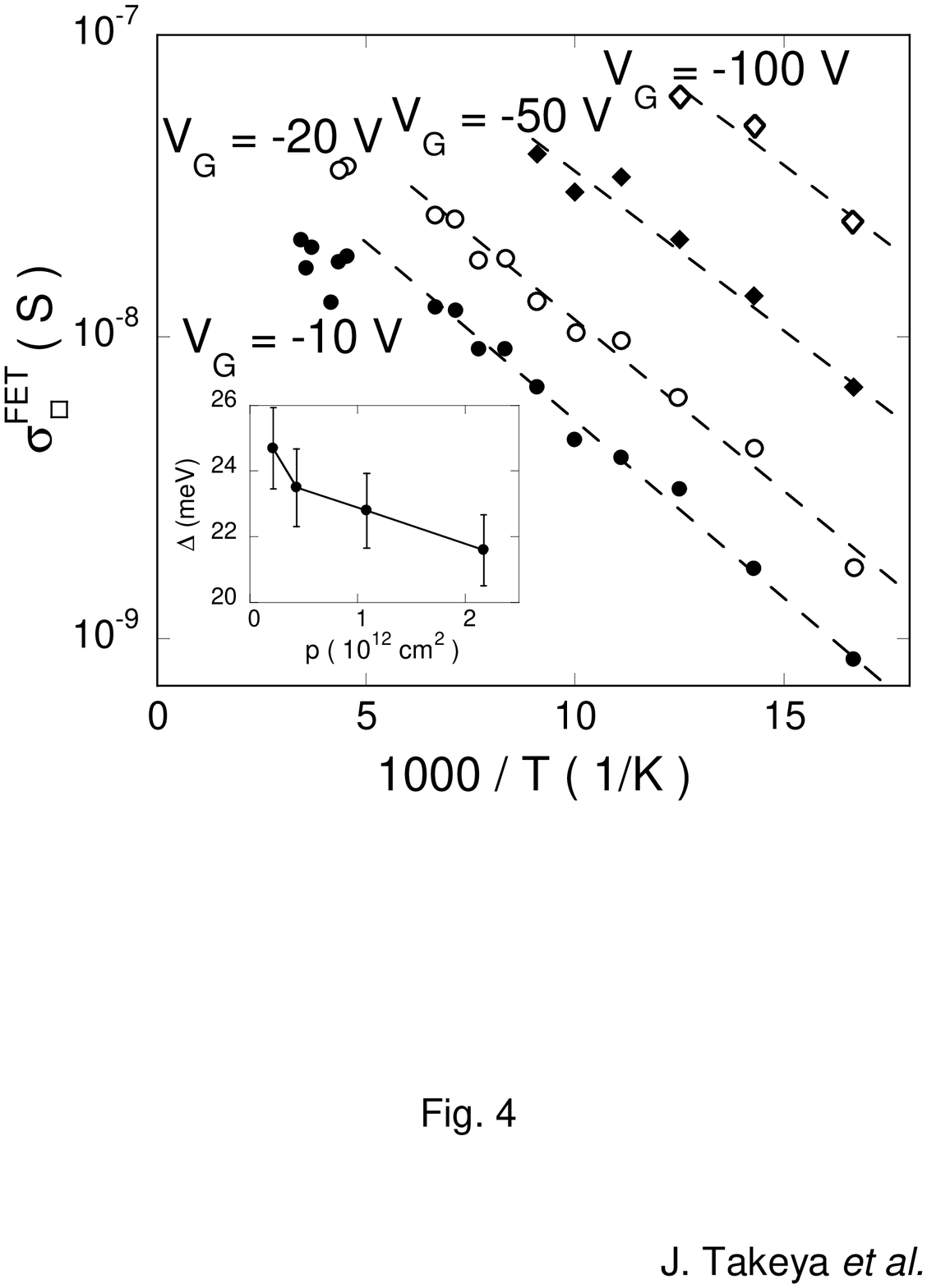}
\end{figure}

\end{document}